\documentstyle[12pt,aaspp4]{article}
\lefthead{Spinrad et al.}
\righthead{$z=5.34$ Galaxy Pair}

\def\cf{{c.f.,~}}
\def\eg{{e.g.,~}}
\def\etal{{et al.~}}
\def\ie{{i.e.,~}}

\def\deg{\ifmmode {^{\circ}}\else {$^\circ$}\fi}

\def\secper{\ifmmode \rlap.{^{s}}\else $\rlap{.}{^{s}} $\fi}

\def\ergsA{\ifmmode {\rm\,erg\,s^{-1}\,\AA^{-1}}\else
    ${\rm\,erg\,s^{-1}\,\AA^{-1}}$\fi}
\def\ergscmA{\ifmmode {\rm\,erg\,s^{-1}\,cm^{-2}\,\AA^{-1}}\else
    ${\rm\,erg\,s^{-1}\,cm^{-2}\,\AA^{-1}}$\fi}
\def\ergcms{\ifmmode {\rm\,erg\,cm^{-2}\,s^{-1}}\else
    ${\rm\,erg\,cm^{-2}\,s^{-1}}$\fi}
\def\kms{\ifmmode {\rm\,km\,s^{-1}}\else
    ${\rm\,km\,s^{-1}}$\fi}
\def\kmsMpc{\ifmmode {\rm\,km\,s^{-1}\,Mpc^{-1}}\else
    ${\rm\,km\,s^{-1}\,Mpc^{-1}}$\fi}

\def\spose#1{\hbox to 0pt{#1\hss}}
\def\simlt{\mathrel{\spose{\lower 3pt\hbox{$\mathchar"218$}}
     \raise 2.0pt\hbox{$\mathchar"13C$}}}
\def\simgt{\mathrel{\spose{\lower 3pt\hbox{$\mathchar"218$}}
     \raise 2.0pt\hbox{$\mathchar"13E$}}}

\def\lya{Ly$\alpha$}
\def\oxytwo{[\ion{O}{2}]$\lambda$3727}

\def\Msun{\ifmmode {\rm\,M_\odot} \else
	${\rm\,M_\odot}$\fi}

\begin{document}

\title{A $z=5.34$ Galaxy Pair in the Hubble Deep Field\altaffilmark{1}}

\author{Hyron Spinrad, Daniel Stern, Andrew Bunker }
\affil{Department of Astronomy, University of California at Berkeley \\
Berkeley, CA 94720 \\
{\tt email: (spinrad,dan,bunker)@bigz.Berkeley.edu}
}
\author{Arjun Dey\altaffilmark{2}}
\affil{Department of Physics \& Astronomy, The Johns Hopkins University \\
3400 N. Charles St., Baltimore, MD 21218 \\
{\tt email: dey@skysrv.pha.jhu.edu}
}
\author{Kenneth Lanzetta, Amos Yahil, Sebastian Pascarelle}
\affil{Department of Physics \& Astronomy, State University of New York at Stony Brook \\
Stony Brook, NY 11794--3800 \\
{\tt email: (lanzetta,ayahil,sam)@sbast3.ess.sunysb.edu}
}
\author{and Alberto Fern\'andez--Soto }
\affil{Department of Astrophysics \& Optics, University of New South Wales \\
Sydney, Australia NSW2052 \\
{\tt email: fsoto@bat.phys.unsw.edu.au}
}

\altaffiltext{1}{Based on observations at the W.M. Keck Observatory,
which is operated as a scientific partnership among the University of
California, the California Institute of Technology, and the National
Aeronautics and Space Administration.  The Observatory was made possible
by the generous financial support of the W.M. Keck Foundation.}
\altaffiltext{2}{Hubble Fellow.}

\begin{abstract}

We present spectrograms of the faint $V$--drop ($V_{606} = 28.1,
I_{814} = 25.6$) galaxy pair HDF~3--951.1 and HDF~3--951.2 obtained at
the Keck~II Telescope.  Fern\'andez--Soto, Lanzetta, \& Yahil (1998)
derive a photometric redshift of $z_{\rm ph} = 5.28^{+0.34}_{-0.41}$
($2\sigma$) for these galaxies; our integrated spectrograms show a large
and abrupt discontinuity near $7710 \pm 5$ \AA.  This break is almost
certainly due to the \lya\, forest as its amplitude ($1 -f_\nu^{\rm
short}/f_\nu^{\rm long} > 0.87, 95\%$ confidence limit) exceeds any
discontinuities observed in stellar or galaxian rest--frame optical
spectra.  The resulting {\it absorption--break} redshift is $z=5.34 \pm
0.01$.  Optical/near--IR photometry from the HDF yields an
exceptionally red ($V_{606}-I_{814}$) color, consistent with this large
break.  A more accurate measure of the continuum depression blueward of
\lya\, utilizing the imaging photometry yields $D_A = 0.88$.

The system as a whole is slightly brighter than $L_{1500}^*$ relative
to the $z \sim 3$ Lyman break population and the total star formation
rate inferred from the UV continuum is $\approx 22\, h_{50}^{-2}\, \Msun\,
yr^{-1}$ ($q_0 = 0.5$) assuming the absence of dust extinction.  The
two individual galaxies are quite small (size scales $\simlt
1 h_{50}^{-1}$ kpc).  Thus these galaxies superficially resemble the
Pascarelle \etal (1996) ``building blocks''; if they comprise a
gravitationally bound system, the pair will likely merge in a time
scale $\sim 100$ Myr.

\end{abstract}


\section{Introduction}

We are presently targeting photometrically--selected faint galaxies for
spectroscopic study at the Keck Telescopes with the goal of measuring
redshifts and star--formation rates at early cosmic epochs.  Selecting
high--redshift galaxies based upon their continuum properties (\cf
Weymann \etal 1998) is important and complements work on emission
line--selected galaxies at $z \simgt 4.5$ found serendipitously (Dey
\etal 1998) and from narrow--band imaging (Cowie \& Hu 1998; Hu, Cowie,
\& McMahon 1998).  Studying galaxies at these high redshifts has
important implications for tracing the formation of galaxies and large
scale structure, mapping the history of star formation, and
understanding the chemical history of the Universe.

The Hubble Deep Field (hereinafter HDF; Williams \etal 1996) has
galvanized a renewed effort at estimating photometric redshifts (\eg
Lanzetta, Yahil, \& Fern\'andez--Soto 1996; Sawicki, Lin, \& Yee 1997;
see also Hogg \etal 1998).  The extremely deep multiband integrations
through the crisp eye of the {\it Hubble Space Telescope} {\it (HST)}
supplemented by several deep campaigns across the electromagnetic
spectrum (\eg at radio, sub-millimeter, far--infrared, and
near--infrared wavelengths --- Fomalont \etal 1997; Hughes \etal 1998;
Rowan--Robinson \etal 1997; Hogg \etal 1997; Eisenhardt \etal 1996;
Thompson \etal 1998) --- are an ideal data set with which to estimate
redshifts based upon broad--band colors.  High--redshift targets are
robustly selected based upon the redshifted Lyman break at $(1 + z)
\times 912$ \AA\ and the redshifted \lya\, discontinuity at $(1 + z)
\times 1216$ \AA, causing the galaxies to effectively disappear, or
``drop out'', at short wavelengths.  $U$--band dropouts, corresponding
to $z \sim 3$, have been systematically studied by several groups (\cf
Steidel \etal 1996a; Lowenthal \etal 1997; Spinrad \etal 1998; Bunker
\etal 1998).  $B$--band dropouts, corresponding to $z \sim 4$, and
$V$--band dropouts, corresponding to $z \sim 5$, are beginning to be
addressed (Dickinson 1998; Weymann \etal 1998; Dey \etal 1998b).  Our
present list of potential $z > 4$ candidates ($B$-- and $V$--dropouts)
includes six galaxies with $I_{814} < 26.5$ in the HDF
(Fern\'andez--Soto, Lanzetta, \& Yahil 1998; AB scale used
throughout\footnote{The AB magnitude system (Oke 1974) is defined such
that $m_{\rm AB}=-2.5\log_{10}(f_\nu\,/\,{\rm erg\,cm}^{-2}\,{\rm
s}^{-1}\,{\rm Hz}^{-1}) - 48.60$.}).  Their spectroscopic study, even
with the large aperture of the Keck Telescopes and dark sky of Mauna
Kea, is clearly a technical challenge.  Weymann \etal's (1998)
confirmation of a galaxy at $z=5.60$ illustrates that the $V$--drop
technique works.  A systematic survey, however, is necessary to assess
that this population is not contaminated by lower--redshift interlopers
such as galaxies with extremely high equivalent width emission lines
(\eg \oxytwo) in the $I_{814}$ filter.  We describe the observations of
one $V$--drop system, HDF~3--951.0.  Our data imply a redshift of $z =
5.34$ for this system, one of the highest redshifts yet measured and
among the first systematically pre--selected galaxies at $z > 5$.

Throughout this paper we adopt $H_0 = 50\, h_{50}\, \kmsMpc, q_0 = 0.5\,
(0.1),$ and $\Lambda = 0$.  For these parameters, 1\arcsec\, subtends
5.6 (10.2) $h_{50}^{-1}$ kpc at $z = 5.34$ and the Universe is only
820 Myr (1.56 Gyr) old, corresponding to a lookback time of 93.7\%
(90.6\%) of the age of the Universe.  We present our observations in the
following section, our redshift determination in \S3, and discuss the
galaxy's inferred properties in \S4.

\section{Observations} 

HDF~3--951.0 is a faint galaxy ``pair'', comprised of HDF~3--951.1 and
HDF~3--951.2, near the edge of the WF3 CCD of the HDF.  In Fig.~1 we
present F606W ($V_{606}$) and F814W ($I_{814}$) images of the galaxy.
Extant photometry of this system is assembled in Table~1.  Clearly the
most outstanding photometric features of the composite (HDF~3--951.0)
energy distribution are the non--detection at $U_{300}$ and $B_{450}$,
the marginal detection in $V_{606}$, and the very red color in $V_{606}
- I_{814}$.  The energy distribution appears to flatten at longer
wavelengths with $I_{814} - K \simlt 2.0$.  These colors qualitatively
suggest a high--redshift, star--forming system with the \lya\, forest
attenuating the spectrum below $I_{814}$ and OB stars dominating the
rest--frame UV past \lya.  A more detailed technique employing template
spectra and maximum likelihood analysis yields a photometric redshift
of $z_{\rm ph} = 5.28^{+0.34}_{-0.41}$ ($2\sigma$) for this system and
$z=5.72^{+0.33}_{-0.34}$ ($2\sigma$) for HDF~3--951.1 alone (the brighter
component; object \#3 in Fern\'andez--Soto, Lanzetta, \& Yahil 1998;
see also Lanzetta, Yahil, \& Fern\'andez--Soto 1996).  Comparisons
between photometric and spectroscopic redshift determinations show that
the former is typically robust to $\Delta z \approx 0.34$ for objects
with $I_{814} \sim 25.5$ and $z > 3$ (Fern\'andez--Soto, Lanzetta, \&
Yahil 1998).

However, an accurate determination of the redshift requires deep
spectroscopy, and so we observed the HDF during three observing runs in
1998 using the spectroscopic mode of the Low Resolution Imaging
Spectrometer (LRIS; Oke \etal 1995) at the Cassegrain focus of the
Keck~II Telescope.  Only the data collected on UT 1998 February 19 were
of high quality; UT 1998 January 20 suffered from poor seeing and high
cirrus, while integrations on UT 1998 March 28 \& 29 were plagued by
poor seeing.   All observations employed milled slitmasks constructed
to allow simultaneous observations of seven $B$-- and $V$--dropout
galaxies.  The 1\farcs5 wide slitlets were typically 20\arcsec\, long,
allowing sufficient slit length for sky subtraction at the expense of a
diminished number of targets.  Slitmask observations were made at a
position angle of $102.6\deg$ (east of north) with the 400 l/mm grating
($\lambda_{\rm blaze} \approx 8500$ \AA; $\Delta \lambda_{\rm FWHM}
\approx 11$ \AA) sampling the wavelength range $\lambda\lambda 5940 -
9720$ \AA.  Small spatial shifts ($\approx 4\arcsec$) were performed
between each $\approx 1800s$ exposure to facilitate removal of fringing
in the near--IR regions of the spectrograms.

All data reductions were performed using the IRAF package and followed
standard slit spectroscopy procedures.  Wavelength calibration was
performed using a NeAr lamp, employing telluric lines to adjust the
zero--point.  Flux calibration was performed using observations of
G191B2B, Feige~34, HZ~44, and Wolf~1346 (Massey \etal 1988; Massey \&
Gronwall 1990) and accurate spectrophotometry was verified against the
HDF photometry by convolving the spectra of HDF~3--951.0 and a brighter
galaxy which serendipitously lay along one of the slitlets
(HDF~3--493.0 --- $I_{814} = 21.74, z=0.848$) with the F814W filter
response function.  HDF~3--951.0 was detected during all three
observing runs.  However, the February data (seeing $\sim$ 0\farcs7,
photometric), comprising four integrations totaling 6900$s$, is of
much higher signal--to--noise ratio; our final spectrum (Fig.~2) is
composed from the February data alone.

\section{Redshift Determination}

Our final spectrogram of the unresolved pair of faint galaxies (Fig.~2)
yields a fairly noisy but robust result:  above 7720 \AA\, there is a
roughly flat continuum (in $f_\nu$)  with a mean flux near $0.4 \mu{\rm
Jy}$.  Almost no light is detected below 7700 \AA\, ($f_\nu < 0.05
\mu{\rm Jy}$), and an accordingly large and abrupt discontinuity exists
at $7710 \pm 5$ \AA.  An accurate measurement of the discontinuity
wavelength is made difficult by the faint magnitudes considered and the
challenges of sky subtraction in the 7700 \AA\, OH sky emission band.
We note that the faintness and 0\farcs61 separation of HDF~3--951.1 and
HDF~3--951.2 makes separate spectroscopy with ground--based
instrumentation exceedingly difficult.  Their unusual, yet similar
colors, however, support the hypothesis that they lie at the same
redshift.  Disparate redshifts would lead to a dilution of the 7710
\AA\, break amplitude.  The spectrum is also obviously inconsistent
with a single high--equivalent width emission line dominating the
$I_{814}$ flux and causing the extremely red $V_{606} - I_{814}$
color.  Our spectrophotometry yields a 2$\sigma$ limit to the equivent
width of an unresolved emission line $W_\lambda^{\rm obs} < 40$\AA\,
for $\lambda > 7800$ \AA.  If the red $V_{606} - I_{814}$ color were
due to an extremely strong emission line in $I_{814}$, the required
equivalent width would be at least $W_\lambda^{\rm obs} > 200$ \AA, for
a constant slope continuum fit to the $V_{606}$ and $K$ upper limit
magnitudes.

Discontinuities of this amplitude ($\simgt 8$) are unprecedented in
optical spectra of stars and galaxies.  Averaging the spectrum in 10
pixel ($\approx 18$ \AA) bins and considering Poissonian counting
statistics, we find the average flux density above the discontinuity is
$f_\nu^{\rm long} (\lambda \lambda 8000 - 9000 {\rm \AA}) = 0.432 \pm
0.052 \mu{\rm Jy}$, while the average flux density below the
discontinuity is $f_\nu^{\rm short} (\lambda \lambda 6500 - 7500 {\rm
\AA}) = -0.036 \pm 0.038 \mu{\rm Jy}$, \ie consistent with no
observable flux.  There is, at this $\approx 28$ magnitude level, a
small systematic problem.  The 95\%\, (99\%) confidence limit to the
amplitude of this continuum depression, calculated from Monte Carlo
simulations of the flux densities with the constraint that $f_\nu >
0$,  is then $1 - f_\nu^{\rm short}/f_\nu^{\rm long} > 0.87$ (0.82).
In order of decreasing wavelength, discontinuities are commonly
observed in UV/optical spectra of galaxies at rest wavelengths of 4000
\AA\, ($D(4000)$), 2900 \AA\, ($B(2900)$), 2640 \AA\, ($B(2640)$), 1216
\AA\, (\lya), and 912 \AA\, (the Lyman limit).  The hydrogen
discontinuities derive from associated and foreground absorption and
thus have no theoretical maximum.  The longer rest--wavelength
discontinuities derive from metal absorption in the stars and galaxies,
and are thus dependent upon the age and metallicity of the galaxy (\cf
Fanelli \etal 1992; Spinrad \etal 1997).  The largest measured values
of $D(4000)$ in lower--redshift, early--type galaxies are $\approx 2.6$
(Dressler \& Gunn 1990; Hamilton 1985), corresponding to $1 -
f_\nu^{\rm short}/f_\nu^{\rm long} \approx 0.62$, while {\it IUE}
spectra of main--sequence stars exhibit $B(2900) \simlt 3$ and $B(2640)
\simlt 3$ (Spinrad \etal 1997), corresponding to $1 - f_\nu^{\rm
short}/f_\nu^{\rm long} \simlt 0.67$.  Therefore, we can safely rule
out the low--redshift interpretations of the spectrum of HDF~3--951.0.
The 7710 \AA\, break is also unlikely to be associated with a Lyman
limit at $z = 7.45$; at that redshift, the \lya\, forest would likely
obliterate the rest--frame $912 - 1216$ \AA\, spectrum.  \lya\, itself
would be at $1.03\mu$m, which is a very challenging wavelength for
current CCD detectors.  Identifying the break with \lya\, is the lowest
redshift and most likely interpretation under these circumstances.

Large breaks are occasionally seen in exotic objects as well.  For
instance, the iron low--ionization broad absorption line quasar (Fe
Lo-BAL) FIRST~J155633.8+351758 (Becker \etal 1997) has a discontinuity
with $1 - f_\nu^{\rm short}/f_\nu^{\rm long} \approx 0.85$ around 2800
\AA.  However, this object belongs to an exceedingly rare type of
quasar, a classification which we can rule out for HDF~3--951.0 due to
its resolved morphology.  Furthermore, radio--loud broad absorption
line quasars tend to have very red optical/near--IR colors (Hall \etal
1997).

We therefore associate the break with the \lya\, forest,  implying a
redshift $z = 5.34 \pm 0.01$.   The systematics of \lya\, absorption
might provide a systematic redward bias of $\Delta z \approx  0.01$ ---
in high--redshift galaxies, associated and foreground absorption
generally displaces \lya\, emission lines redward of their host galaxy
systematic velocity by up to several hundred \kms.  In fact, this
mechanism also imprints an asymmetry onto the \lya\, emission line,
when present, thus providing a powerful discriminant between
high--redshift \lya\, and low--redshift \oxytwo\, (see Dey \etal
1998a).  The sharpness of the discontinuity and the flatness of the
longer--wavelength spectrogram (in $f_\nu$) are further arguments for
identifying the break with the \lya\, forest onset.  For the remainder
of this paper we adopt $z=5.34$ for the redshift of the faint galaxy
pair HDF~3--951.0.

\section{Discussion}

Our deep spectroscopy confirming the high--redshift of HDF~3--951.0, as
well as Weymann \etal's (1998) confirmation of HDF~4--473.0 at
$z=5.60$, illustrates that the photometric redshift technique, and, in
particular, $V$--drop selection, is a robust method for selecting and
studying the distant Universe.  We now derive some basic physical
properties of HDF~3--951.0, consonant with its faint magnitude.

If the UV continuum is dominated by light from young, hot stars, the
star--formation rate ($\dot{M}$) may be derived from the UV flux at
$\lambda_0 \simgt 1240$ \AA.  Assuming the continuum emission from
HDF~3--951.0 is unreddened and has a spectral slope of $f_\nu \propto
\lambda^0$, consistent with the observations, we derive $L_{1500} =
22.9 \times 10^{40}\, h_{50}^{-2}\, \ergsA$ and $M_{1500} = -21.5$ AB
mag for HDF~3--951.0 based upon the flux density between 8000 and 9000
\AA\, (see Table~1).  Madau, Pozzetti, \& Dickinson (1998) calculate
$\dot{M} \approx 10^{-40}\, L_{1500} \Msun\, yr^{-1}$ for $L_{1500}$
measured in units of \ergsA and a $> 100$ Myr old population with a
Salpeter IMF ($0.1 < M < 125 \Msun$).  This is roughly consistent with the
relation derived from the Leitherer \& Heckman (1995) models for a
different IMF and much younger ages of $< 10$ Myr.  These conversions
are meant to be illustrative rather than definitive; they depend upon
the assumed star--formation history, IMF, metallicity, and age.  The
lower limit on the inferred star--formation rate for HDF~3--951.0 is
thus $\approx 22\, h_{50}^{-2}\, \Msun\, yr^{-1}$, assuming the absence
of dust absorption.  Dickinson (1998) finds that the ultraviolet
luminosity function of Lyman--break galaxies at $z \approx 3$ is
well--modeled by a Schechter luminosity function of characteristic
absolute magnitude $M_{1500}^* \approx -21$ AB mag.  This implies that
the HDF~3--951.0 galaxies are individually sub--luminous, but slightly
brighter than $L_{1500}^*$ when considered as a single system
uncorrected for extinction.

The star--formation rates may also be determined for the galaxy pair
individually utilizing the imaging photometry, without reference to the
spectrophotometry.  Assuming a Heaviside function spectrum with $f_\nu
= 0$ below 7710 \AA, and $f_\nu \propto \lambda^0$ redward of 7710 \AA,
the $I_{814}$ magnitudes may be used to calculate an upper limit on the
flux density redward of \lya.   This then yields $F_{1500}$ and, with
the above prescription, the inferred star formation rate (see
Table~1).  We find $\dot{M} = 13 h_{50}^{-2}\, \Msun\, yr^{-1}$ for
HDF~3--951.1 and $\dot{M} = 6 h_{50}^{-2}\, \Msun\, yr^{-1}$ for
HDF~3--951.2.  

The flat red end of the spectrum of HDF~3--951.0 is similar to spectra
of $z \sim 3$ Lyman break galaxies (\cf Steidel \etal 1996a) ---
systems which are well--represented by an OB stellar population with
little dust.  Deep Keck/LRIS spectroscopy of some of the brighter $z
\sim 3$ Lyman break population suggests that galaxies with \lya\, in
emission are generally flatter in $f_\nu$ at $\lambda \lambda 1220 -
1700$ \AA, while those galaxies with \lya\, in absorption are generally
redder at these wavelengths (Spinrad \etal 1998).  From a study of
vacuum--UV {\it IUE} spectra of local starburst galaxies, Heckman \etal
(1998) find that metal--rich starbursts are redder and more heavily
extinguished (have larger values of $L_{\rm IR}/L_{\rm UV}$), have
stronger rest--frame UV absorption lines, and occur in more massive and
brighter host galaxies.  Similarly, the brightest Lyman break galaxies
tend {\em not} to have \lya\, in emission (Steidel, private
communication), possibly a result of the galaxies lying in deeper
potential wells and thus being more able to retain gas and dust which
scatter and absorb the \lya\, photons and redden the $\lambda > 1220$
\AA\, continuum.  In this scenario, the apparent flatness of our HDF~3--951.0
spectrum at $\lambda_0 > 1300$ \AA\, is inconsistent with the lack of a
measurable \lya\, emission line.  However, relatively small column
densities of neutral gas with even very small dust content can destroy
\lya\, emission if this gas is static with respect to the ionized
region where \lya\, photons originate (\cf Kunth \etal 1998).

HDF~3--951.0 is potentially reddened by foreground and associated dust,
consistent with the lack of \lya\, emission; our ground--based limits
on the near--IR magnitude of the system constrains $E_{\rm B-V} \simlt
0.3$ (for a dust--free Heaviside spectrum subject to extinction by a
foreground screen of dust following the extinction law of Cardelli,
Clayton, \& Mathis 1989).  This level of extinction would imply
intrinsic star formation rates $\sim 45$\% higher than the values
quoted above.  A real measurement of the dustiness of the galaxy pair
must await deep near--IR images of the field, as have recently been
obtained with NICMOS on the {\it HST}.

We next consider the $I_{814}$ morphology of this system (see Fig.~3).
The full--width, half--maxima (FWHM) of HDF~3--951.1 and HDF~3--951.2
are 0\farcs50, and 0\farcs28 respectively.  Comparison with a star
reveals that both are clearly resolved (FWHM$_{\rm star} = 0\farcs14$)
with deconvolved half--width, half--maxima (HWHM) of 0\farcs24 and
0\farcs12 respectively.  For $q_0$ = 0.5 (0.1),
these correspond to 1.3 (2.5) $h_{50}^{-1}$ kpc for HDF~3--951.1 and
0.7 (1.2) $h_{50}^{-1}$ kpc for HDF~3--951.2, comparable to the values
found for many of the $z \approx 3$ Lyman--break galaxies (Giavalisco,
Steidel, \& Macchetto 1996).  HDF~3--951.1 (the brighter component)
contains sub--structure, with a second ``hot spot'' $\sim$~0\farcs12
east of the core, at a projected separation of 0.66 (1.2) $h_{50}^{-1}$
kpc.  We speculate that this is either a knot of star formation (bright
in the rest--frame UV), or evidence of multiple nuclei.  The projected
proximity of HDF~3--951.2 adds weight to the hypothesis that this is a
dynamically--bound system, and that we are witnessing a merger event.
Lyman--break galaxies at $z\approx 3$ often exhibit either disrupted
morphologies or multiple components (\eg Giavalisco, Steidel, \&
Macchetto 1996; Steidel \etal 1996b; Bunker \etal 1998).

Due to the sub--structure of its core, HDF~3--951.1 is not fit well by
either a de~Vaucouleurs $r^{1/4}$ law nor an exponential surface
brightness profile. The exponential disk appears to dominate in a
two--component model, and the scale length is 0\farcs23, equivalent to
$r_{\rm disk}$ = 1.3 (2.3) $h_{50}^{-1}$ kpc. The elongation is
$b/a=1.2$.

The fainter HDF~3--951.2 is well--fit by an exponential disk profile,
with a scale length of 0\farcs12, corresponding to 0.65 (1.2)
$h_{50}^{-1}$ kpc, and is almost circular ($b/a=1.1$). As with the
$z\approx 3$ population, we note that HDF~3--951 is significantly more
compact at rest--frame UV wavelengths compared to local disk galaxies,
which have typical scale lengths of $\sim$~5~kpc at optical wavelengths
(Freeman 1970).

The angular separation of 0\farcs61 projects to 3.4 (6.2) $h_{50}^{-1}$
kpc for $q_0 = 0.5\, (0.1)$, implying that HDF~3--951.0 is a pair of
sub--luminous systems of modest projected separation.  What can we say
about the evolutionary fate of HDF~3--951.0?  Given the small physical
sizes and projected separation, in all likelihood HDF~3--951.1 and
HDF~3--951.2 will merge into a single galaxy.  Assuming a relative
velocity of $\Delta v = 200$ km s$^{-1}$ and a physical separation
equal to the projected $\approx 5$ kpc, the crossing time is $\approx
25$~Myr.  Thus we estimate the merger time scale for HDF~3--951.1 and
HDF~3--951.2 is a few crossing times (\cf Barnes \& Hernquist 1996), or
$\sim 100$ Myr.  Indeed, we suggest that HDF~3--951.1 and HDF~3--951.2
are already in the process of merging; we are perhaps witnessing the
galaxies in a post--collision state, with the luminosity enhanced by
merger--induced star formation.  Studies of low--redshift merging
systems find enhanced rates of star formation (Sanders \etal 1988),
consistent with the apparently OB star--dominated spectrum of
HDF~3--951.0.  We note that our $I_{814}$ images sample the rest--frame
UV.  Even in present--day galaxies, UV--emitting regions in galaxies
are typically small.  Alternatively, two regions of active
star--formation within $\approx$ 5 kpc of each other may well be
star--forming knots within the same galaxy.  Longer--wavelength imaging
will help resolve this question.  Indeed, preliminary reductions of
NICMOS observations of the HDF suggest that these objects remain
separate in F160W images (rest--frame $\sim 2500$ \AA; Dickinson,
private communication).  We note, finally, that the size and luminosity
of HDF~3--951.0 suggest it to be a more distant version of the $z
\approx 2.4$ galaxies discussed by Pascarelle \etal (1996) which have
typical half--light radii of $0\farcs1 - 0\farcs2$.

What are the implications of the large continuum discontinuities in the
\lya\, region?  How reliable is the nomimal factor of 10 at 95\%\,
confidence level we suggest, and how does it propagate to the $D_A$
parameter (Oke \& Korycansky 1982, Madau 1995, Schneider \etal
1991ab)?  The largest discontinuities previously measured in quasar
spectra at $z \simgt 4.5$ are approximately a factor of 4 (see
Fig.~4).  Our determination of the $f_\nu$ break amplitude for
HDF~3--951.0 is made difficult since the composite galaxy spectrum is
very faint at $\lambda \lambda 6500 - 7700$ \AA.  Oke \& Korycansky
(1982) define $D_A$ as $$D_A \equiv \langle 1 - {{f_\nu(\lambda\lambda
1050 - 1170)_{\rm obs}}\over {f_\nu(\lambda\lambda 1050 - 1170)_{\rm
pred}}}\rangle.$$ From the February 1998 data, we measure average flux
densities of $f_\nu(\lambda \lambda 1050 - 1170) = -0.029 \pm 0.045
\mu{\rm Jy}$ and $f_\nu(\lambda \lambda 1250 - 1370) = 0.405 \pm 0.055
\mu{\rm Jy}$. The implied 95\% (99\%) confidence limit to $D_A$ is then
$D_A >$ 0.82 (0.75).  The 1998 January and March data qualitatively
support our large amplitude break, but quantitatively do not aid our
numerical evaluation of it.  Larger values are possible and future,
more sensitive observations will determine a more precise value of
$D_A$ for this system.

Alternatively, we can produce a photometric estimate of $D_A$ utilizing
the Williams \etal (1996) broad band colors, the {\it HST}/WFPC2 filter
curves, and the plausible assumption (born out by our spectra to date)
that the $f_\nu$ flux distributions above and below \lya\, are flat.
We assume a two--step spectral energy distribution with zero flux below
the Lyman limit, $f_\nu^-$ between the Lyman limit and \lya, and
$f_\nu^+$ redwards of \lya.  For $f_\nu^\pm \propto \lambda^0$, 54\% of
the F606W flux comes from $\lambda > 5782$ \AA\, (the Lyman limit at
$z=5.34$) and 48\% of the F814W flux comes from $\lambda > 7710$ \AA\,
(\lya\, at $z=5.34$).  For the observed magnitudes of HDF~3--951.0,
this implies $f_\nu^- = 0.04 \mu$Jy and $f_\nu^+ = 0.33 \mu$Jy.  The
resultant ``photometric \lya discontinuity'', with $V - I = 2.50$ is
$D_A^{\rm phot} = 0.88$.  This agrees with the break amplitude derived
from the Keck spectrogram and is much higher than previously reported
values of $D_A$ (also see Dey \etal 1998).  In Fig.~4 we present recent
measurements of $D_A$ from spectra of quasars and high--redshift
galaxies, where HDF~3--951.0 is indicated by the more robust
photometric measurement.  The Weymann \etal (1998) points utilize their
photometry, corrected for the \lya\, emission line flux of $1.0 \times
10^{-17}$ \ergcms, and a $f_\nu \propto \lambda^\beta$ spectrum fit to
the NICMOS near--IR brightnesses (top point).  The lower Weymann \etal
(1998) point utilizes an $f_\nu \propto \lambda^{-0.4}$ spectrum,
corresponding to their best--fit semi--empirical model.

Our concern about the break amplitude arises from its strength:
Madau's (1995) theoretical estimate of the contribution of the \lya\,
forest to $D_A$ is only $\approx$ 0.79 at $z=5.34$ and $\approx$ 0.83
at $z=5.60$.  This extrapolation assumes a distribution of high and low
optical depth foreground \lya\, clouds causing Lyman series absorption
in the spectrum of a distant quasar or galaxy.  The scatter around the
Madau curve is substantial, even at lower redshifts, so the high values
of $D_A$ at $z > 5$ may simply reflect the usual scatter observed in
that parameter.  However, at large enough redshift our line of sight
{\em must} penetrate the end stages of reionization (\cf Loeb 1998;
Miralda--Escud\'e \& Rees 1997) where a smooth distribution of neutral
hydrogen gas will cause an additional Gunn--Peterson \ion{H}{1} opacity
at $\lambda < 1216$ \AA.  Whether this starts at $z = 5$ or $z \simgt
10$ remains an intriguing question.  Are we seeing the first hints of
the Gunn--Peterson trough in these distant systems?  If we see enhanced
(Gunn--Peterson) absorption short--ward of \lya\, (corresponding to $z
\sim 5$) relative to the expected thickening of the \lya\, forest, one
might begin to further exploit spectrophotometry of these distant
galaxies in a novel and useful manner.

\acknowledgments

We thank J. Aycock, W. Wack, R. Quick, T. Stickel, G. Punawai, R.
Goodrich, R. Campbell, T. Bida, and B. Schaeffer for their invaluable
assistance during our observing runs at the W.M.  Keck Observatory.  We
are grateful to A. Philips for providing software and assistance in
slitmask construction and alignment and J. Cohen for supporting LRIS;
and to M. Dickinson, E. Gawiser, J.R. Graham, C. Manning, F. Marleau,
and  C. Steidel for useful comments.  We also thank the referee, D.W.
Hogg, for timely and constructive comments.  H.S. acknowledges support
from NSF grant AST~95--28536, D.S. acknowledges support from IGPP
grant 99--AP026, A.B. acknowledges support from a NICMOS postdoctoral
fellowship, A.D. acknowledges support from NASA grant
HF--01089.01--97A, K.L. acknowledges support from NASA grant NAGW--4422
and NSF grant AST--9624216, A.F.--S. acknowledges support from an
Australian ARC grant.


\begin{deluxetable}{lcccccc}
\tablewidth{0pt}
\tablecaption{Photometry.}
\tablehead{
\colhead{Galaxy Component}  & \colhead{$V_{606}$} &
\colhead{$I_{814}$}         & \colhead{$K$} &
\colhead{$\langle F_{1500} \rangle$ \tablenotemark{\S}} &
\colhead{$L_{1500}\,h_{50}^2$ \tablenotemark{\dag}} &
\colhead{$\dot{M}\,h_{50}^2$ \tablenotemark{\ddag}}}
\startdata
HDF~3--951.1 (brighter) & 28.68 & 26.20 & \nodata & 8.0 & 
12.7 & 13 \nl
HDF~3--951.2 (fainter)  & 28.87 & 26.95 & \nodata & 4.0 & 
 6.3 &  6 \nl
\nl
HDF~3--951.0 (sum)      & 28.08 & 25.60 & $>23.6$ &
13.9 (14.4)\tablenotemark{\dag\dag} & 
22.1 (22.9)\tablenotemark{\dag\dag} & 22 (23)\tablenotemark{\dag\dag} \nl
\enddata

\tablecomments{All magnitudes are in the AB system.  Separation of
component centers is 0\farcs61.  Optical isophotal magnitudes are from
Williams \etal (1996), 2$\sigma$ limit on the  near--IR magnitude is
derived from ground--based observations with IRIM (Eisenhardt \etal
1996) for a bidimensional Gaussian with FWHM $\approx 1\farcs20$.
HDF~3--951 is undetected in $U_{300}$ and $B_{450}$, implying 2$\sigma$
limiting magnitudes of $U_{300} > 28.2$ and $B_{450} > 28.9$.  The
small inconsistency in that $I_{\rm HDF~3-951.0} < I_{\rm HDF~3-951.1}
+ I_{\rm HDF~3-951.2}$ is present in Williams \etal (1996) and likely
derives from the faint magnitudes and close separations considered.}

\tablenotetext{\S}{Flux density at 1500 \AA, $F_{1500}$, is in units of
$10^{-20} \ergscmA$ and is derived from $I_{814}$ assuming a $f_\nu
\propto \lambda^0$ spectrum for $\lambda > 7710$ \AA\, with no flux
below 7710 \AA.  See text for details.}

\tablenotetext{\dag\dag}{First value derives from photometry, second
(parenthetical) value derives from spectrophotometry utilizing the
continuum flux near 1350 \AA\, and assuming a $f_\nu \propto
\lambda^0$ spectrum.  See text for details.}

\tablenotetext{\dag}{$L_{1500}$ is in units of $10^{40} \ergsA$, calculated for
$q_0 = 0.5$ using $F_{1500}$ values.}

\tablenotetext{\ddag}{Star--formation rates, in units of $\Msun\, yr^{-1}$,
assume $q_0 = 0.5$ and a Salpeter IMF with $0.1 < M < 125 \Msun$ (see
Madau, Pozzetti, \& Dickinson 1998 for details).  For $q_0 = 0.1$ these
rates are $\approx 3.3$ times larger.}

\label{photometry}
\end{deluxetable}



\begin{figure}
\figurenum{1}
\plotone{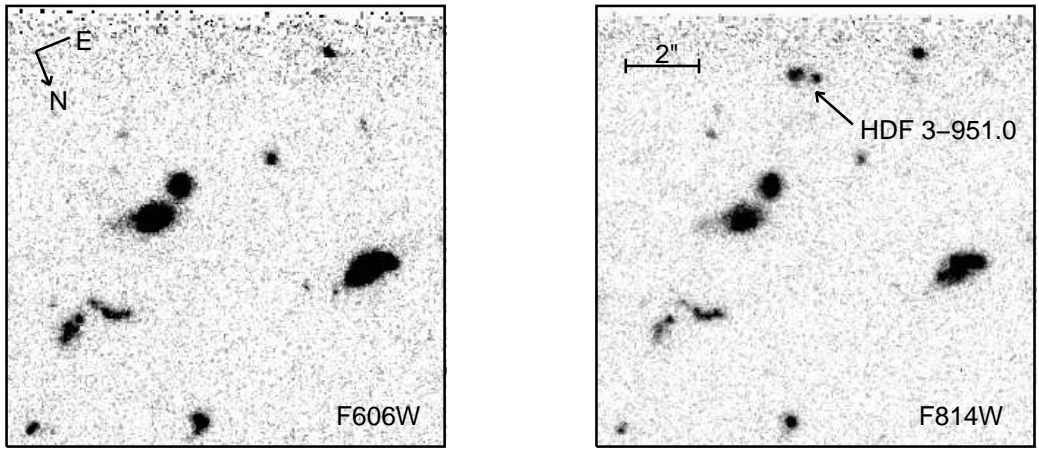}

\caption{{\it HST} F606W (left) and F814W (right) drizzled images of
HDF~3--951.0, a $z=5.34$ galaxy pair comprised of HDF~3--951.1 (SW,
brighter) and HDF~3--951.2 (NE, fainter).  HDF~3--951.0 is located at}

\label{hdfimage}
\end{figure}


\begin{figure}
\figurenum{2}
\plotone{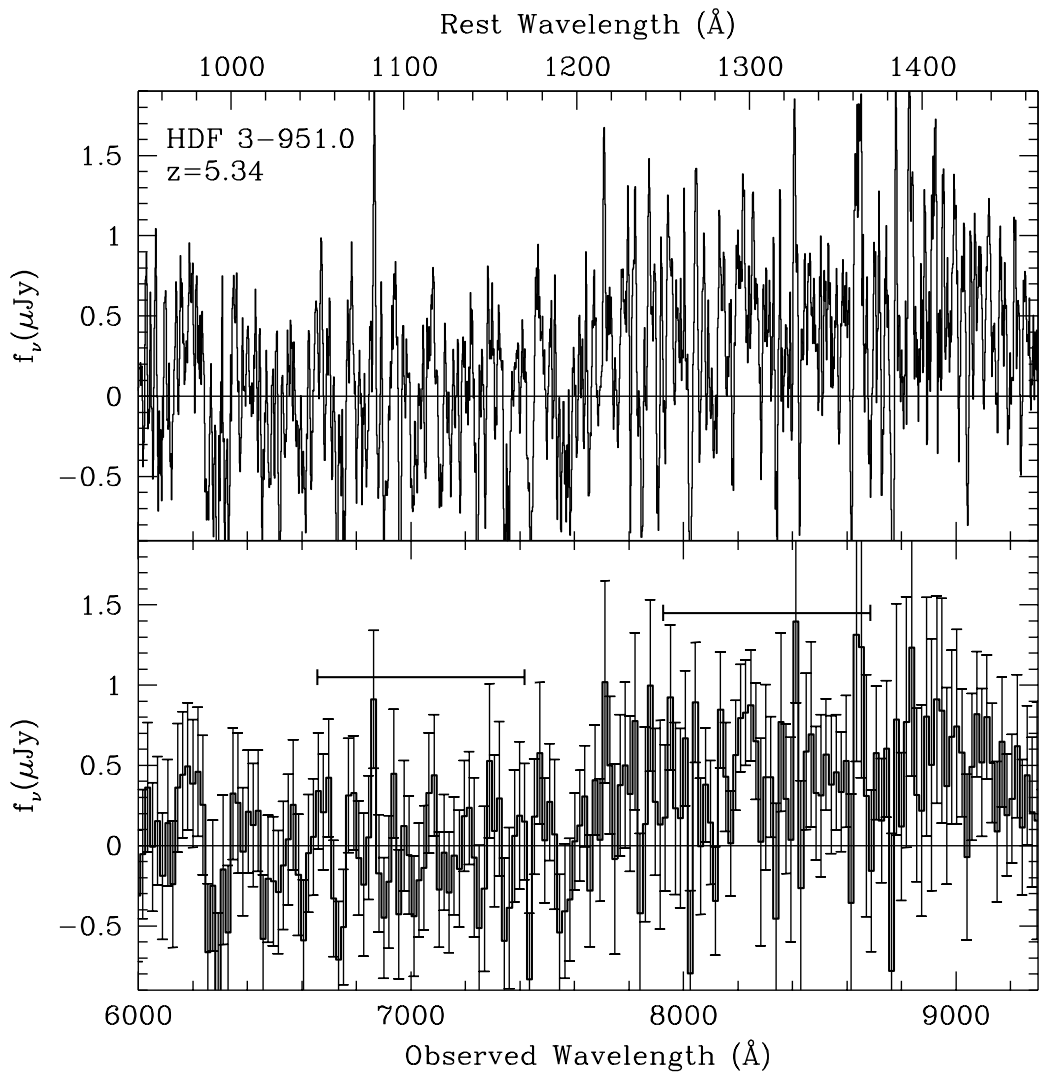}

\caption{Spectrum of the color--selected galaxy HDF~3--951.0 at
$z=5.34$.  Top spectrum is smoothed with a 5 pixel boxcar filter,
bottom spectrum is co--averaged in 10 pixel bins with 1$\sigma$ error
bars assigned according to sky counts.  The total exposure time is
6900$s$, and the spectrum was extracted using an 1\farcs3 $\times$
1\farcs5 aperture.  Horizontal bars on the bottom panel indicates
the wavelength region considered for determination of $D_A$.}

\label{hdfspec}
\end{figure}


\begin{figure}
\figurenum{3}
\plotone{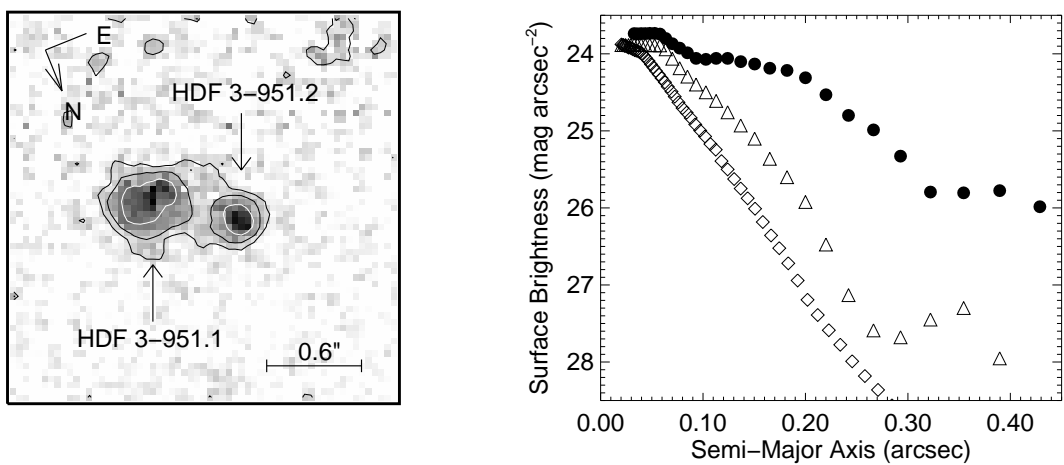}

\caption{Detail of the drizzled F814W ($I_{814}$) image of HDF~3--951.0
(left) and surface brightness profiles for the individual components
(right).  The scaled surface brightness profile of a star is
illustrated (open diamonds); both components of HDF~3--951.0 are
clearly resolved.  Note that the brighter component (HDF~3--951.1;
solid circles) has a sub--structure to the east, possibly indicative of
a recent or ongoing interaction.  The surface brightness profile
clearly illustrates this sub--structure.  The fainter component
(HDF~3--951.2; open triangles) is well--fit by an exponential disk
profile.  The flatness of the surface brightness profiles at small
($\simlt 0\farcs05$) radii are due to sampling the same pixel, and at
large radii ($\simgt 0\farcs3$), the profiles of the galaxies are
contaminated by their respective neighbors.}

\label{morphology}
\end{figure}


\begin{figure}
\figurenum{4}
\plotone{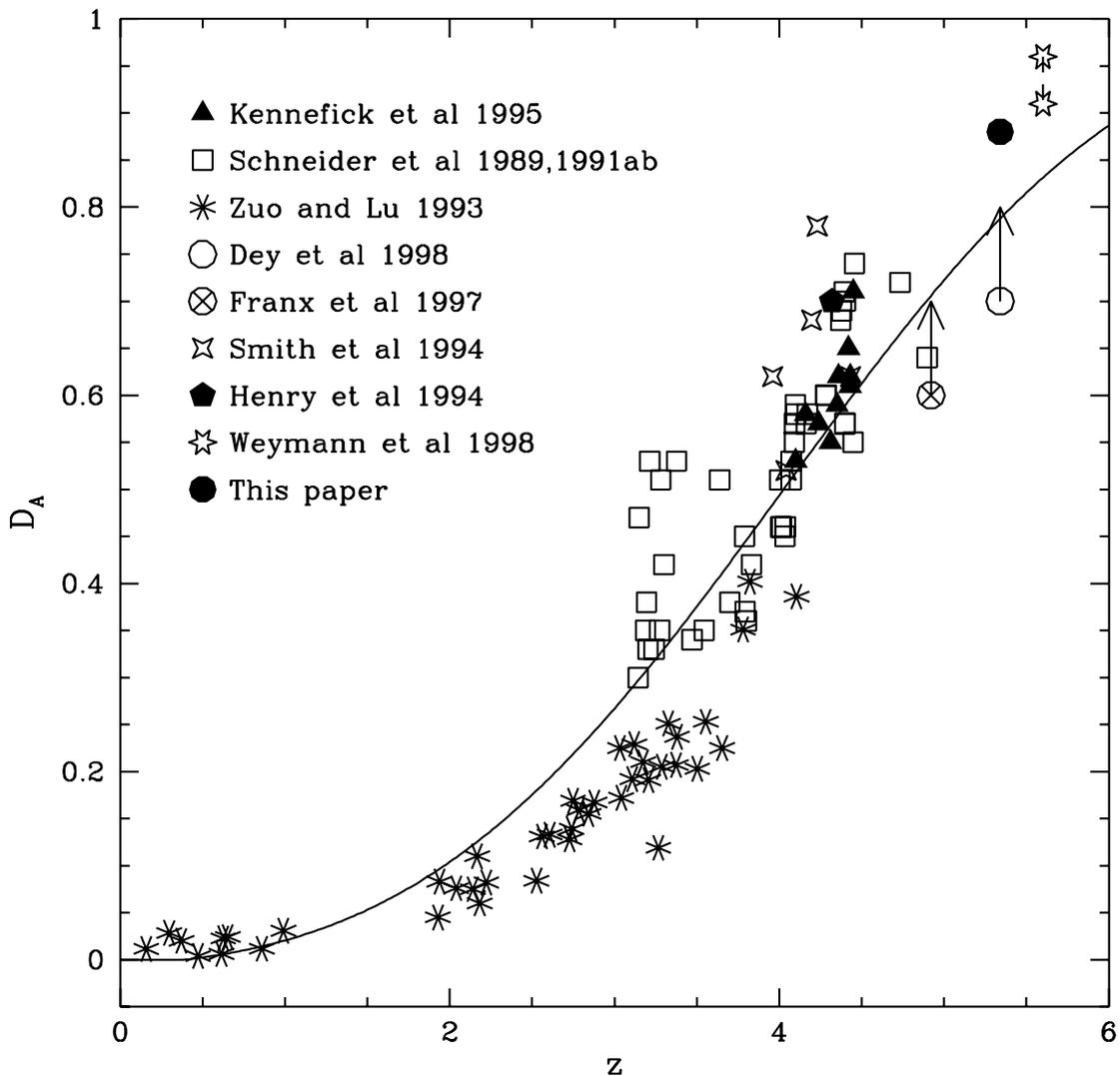}

\caption{Values of the continuum depression blueward of \lya\, ($D_A$)
plotted as a function of redshift, with the Madau (1995) model
overplotted as a solid line.  See text for details regarding derivation
of $D_A$ for HDF~3--951.0 (this paper; photometric measurement) and
HDF~4--473.0 (Weymann \etal 1998).  The apparent systematic
displacement of the Zuo \& Lu (1994) points likely derives from their
revised approach for determining the continuum blueward of \lya:
employing high signal--to--noise, high resolution spectra, they model
and replace the \lya\, forest absorption features.}

\label{da}
\end{figure}

\end{document}